\documentclass[11pt]{article}

\usepackage{
graphicx,amsmath,amssymb,bm}

\newtheorem{prop}{Prop}[section]
\newtheorem{lemma}[prop]{Lemma}
\newtheorem{definition}[prop]{Definition}
\newtheorem{theorem}[prop]{Theorem}

\title{Absence of replica symmetry breaking \\
in disordered FKG-Ising models under uniform field
 }

\author{C. Itoi \ and Y. Utsunomiya  \\
Department of Physics, GS and CST, Nihon University, \\
Kanda-Surugadai, Chiyoda, Tokyo 101-8308, Japan} 
\begin{document}
\maketitle
\begin{abstract} 
We prove that the variance of spin overlap vanishes in disordered Ising models
 satisfying the Fortuin-Kasteleyn-Ginibre (FKG) inequality under a uniform field, such as generally distributed random field Ising model,
 site- and bond-diluted Ising models with the Bernoulli distribution.     
 Chatterjee's proof for the  Gaussian random field Ising model
  is generalized to other  independent identically distributed quenched disorder under a uniform field.   
 \end{abstract}
\section{Introduction}
Replica symmetry breaking is a spontaneous symmetry breaking phenomenon in spin models with quenched disorders.
This symmetry breaking appears generally in mean field disordered  spin models at low temperature, such as
Parisi's replica symmetry breaking formula \cite{Pr,SK} for the Sherrington-Kirkpatrick (SK) model  rigorously proven by Talagrand \cite{T2}.
 Krzakala,  Ricci-Tersenghi, Sherrington and Zdeborova have
 pointed out an evidence that an extended spin glass phase does not exist in disordered ferromagnetic spin models satisfying  the  Fortuin-Kasteleyn-Ginibre (FKG) inequality \cite{FKG}, such as 
 the random field Ising model, diluted Ising model,  the random field Ginzburg-Lamdau model and  random temperature Ginzburg-Landau model
  \cite{K,K2}.
Recently, Chatterjee has proven  that there is no replica symmetry breaking phase in the random field Ising model in an arbitrary dimension rigorously, if the distribution of random field
is Gaussian \cite{C2}.
He  has proven that the variance of the spin overlap vanishes for almost all coupling constants in the random field Ising model, which
satisfies  the FKG inequality \cite{FKG} and the Ghirlanda-Guerra identities.  There have been several studies to generalize Chatterjee's proof and some mathematical results to
non-Gaussian distributions of disorders or quantum spin systems.  
His method has been generalized to quantum systems having the weak FKG property \cite{I3}.  
Auffinger and W.K. Chen have argued disordered spin systems with  generalized distribution and have proven that the overlap is self-averaging in the random field Ising model with a weak coupling constant depending on the system size. They also have proven the Ghirlanda-Guerra identities
and  the  ultrametricity in the mixed $p$-spin model with generalized disorder,  and has shown disorder and temperature chaos phenomena in both the mixed $p$-spin model and the Edwards-Anderson (EA) model  with non-Gaussian distributions.
Roldan and Vira have proven the absence of replica symmetry breaking for
 a different  specific class of non-Gaussian distributions in the random field Ising model \cite{RV}.  Their proof needs  the decay of 
 random fields far from the origin of the lattice.
Quite recently,  Y.T. Chen has improved the arguments  by Auffinger and W. K.  Chen \cite{AC} for the $p$-spin models with a generalized disorder  \cite{CY}.
Barbier, Chan and Macris have proven the concentration of multi-overlaps for random ferromagnetic spin models by a non-trivial application of the Griffiths-Kelly-Sherman
correlation inequality \cite{BCM}.  
The absence of replica symmetry breaking has been  proven  also in the random field Ginzburg-Landau model with Gaussian random field
\cite{IU}. 
In  the present paper, we prove that the replica symmetry breaking does not occur in disordered  Ising systems under general  independent identically distributed (i.i.d.) random fields
satisfying the FKG inequality, such as the
random field Ising model and bond-diluted Ising model under uniform field. The 
 concentration of the multi-overlap obtained by Barbier, Chan and Macris \cite{BCM}
for the multiplicity $k=2$ is similar to our result  but a bit weaker than ours,
since we  prove it without  an assumption of  the Griffiths-Kelly-Sherman
correlation inequality \cite{Gff1,Gff2,KS}.
To prove it for such models,  we employ several new methods 
 as well as that used in Ref\cite{I4,I5}.
\section{Definitions and main theorem}
\subsection{Hamiltonian of models} 
Let $L$ be a positive integer, and define a $d$-dimensional hyper cubic lattice by $\Lambda_L:= [1,L]^d \cap {\mathbb Z}^d$ 
whose volume is $|\Lambda_L|=L^d$.  Let $B_L:=\{ \{x,y\} | x,y \in \Lambda_L, C(x,y) \}$
 be a collection of interaction bonds, where $C(x,y)$ is a condition on two sites $x,y\in \Lambda_L$.
 Assume that the interaction is short-ranged and $|B_L | = C_1 |\Lambda_L|$ for  a certain constant  $C_1 \geq 1$  independent of $L$.
For example, $C(x,y)$ is given by $|x-y|=1 $ to define the collection of  nearest neighbor  bonds  in  $\Lambda_L$. In this case, $C_1=d$.
Let $C_L$ be a collection of interaction ranges defined by $C_L:=B_L$ or $C_L:= \Lambda_L$  to construct FKG-Ising systems.
Note that   $|C_L| \leq C_1 |\Lambda_L|$.
To define Hamiltonian, we introduce random couplings.
Let   $r:=(r_i)_{i\in C_L}$ be real valued i.i.d. random variables with 
a finite expectation and a finite variance.  
Let  $J:=(J_X(r) )_{X\in B_L}$ be a sequence of bond random variables  which 
consist of  positive semi-definite valued functions of the i.i.d. random variables  
$r=(r_i)_{i\in C_L}$, such that  $J_X(r)$ and $J_Y(r)$ are independent if 
 $X \cap Y =\phi$ for two sets $X,Y \in C_L$. 
 Let $(h_x(r) )_{x \in \Lambda_L}$ be a sequence of site random variables which 
consist of  functions of i.i.d. random variables  
$r=(r_i)_{i\in C_L}$, such that  $h_x(r)$ and $h_y(r)$ are independent if $x\neq y$.
Define Hamiltonian as a function of  the sequence $r$ of  random variables 
and spin configurations $\sigma
\in \Sigma_L := \{-1,1\}^{\Lambda_L }$ on 
the lattice $\Lambda_L$. 
 \begin{equation}
 H(r,\sigma) = -\sum_{X \in B_L} J_X(r) \sigma_X - \sum_{x\in \Lambda_L} h_x(r) \sigma_x.
 \label{hamil}
 \end{equation}
 
 \subsection{Examples}
 Here we introduce several examples in our definition of the Hamiltonians in general models. 
  \paragraph{1. Random field Ising model}
This model is defined by  deterministic bond couplings $J_X(r)=1$ for any $X \in B_L$ and site couplings $h_x(r)=b r_x +h$ for any 
 $x \in C_L:= \Lambda_L$ for $(b,h) \in {\mathbb R}^2$, where
 $|\Lambda_L|= L^d$.
 \begin{equation}
 H_{\rm RFI}(r,\sigma) = -\sum_{X \in B_L}  \sigma_X - \sum_{x \in \Lambda_L} (b r_x +h) \sigma_x,
 \end{equation}
 where each random variable $r_x$ at $x \in\Lambda_L$ satisfies a certain distribution with zero expectation and 
 a finite variance.
 \paragraph{2. Bond-diluted Ising model}
 This model is defined by random bond coupling 
 $J_X(r) =J  r_X$  for $J>0$ and deterministic site coupling $h_x(r)=h \in{\mathbb R}$ for $X \in C_L:=B_L$
  \begin{equation}
 H_{\rm BDI}(r,\sigma) = -J  \sum_{X \in B_L} r_X \sigma_X - \sum_{x\in \Lambda_L} h \sigma_x,
 \end{equation}
 where the random variable $r_X$ for $X\in B_L$ satisfies the Bernoulli distribution for $0< p <1$
 \begin{equation}
 p(r_X) = p \delta(r_X-1) + (1-p) \delta (r_X).
 \end{equation}
 Note that $J_X(r)$ and $J_Y(r)$ are independent if and only if $X\neq Y$.
The variance of each $r_X$ is $v=p(1-p)$ in this distribution.
 \paragraph{3. Site-diluted Ising model}  
Spins are missing at several sites in the site-diluted Ising model.
 This model 
 has random bond coupling $J_X(r) =J \prod_{x\in X} r_x$  for $X\in B_L$
 and random site coupling $h_x=h r_x$ for $x\in C_L:=\Lambda_L$ with $h \in{\mathbb R}$ , where $|\Lambda_L|=L^d$. In this model,
 $J_X(r)$ and $J_Y(r)$ are independent if and only if $X \cap Y = \phi$.
 The Hamiltonian of the site-diluted Ising model is defined by
 \begin{equation}
H_{\rm SDI}(r,\sigma) = -J \sum_{X \in B_L}r_X \sigma_X - \sum_{x\in \Lambda_L} h r_x \sigma_x,
\end{equation}
 where the the random variable $r_x$ at $x\in \Lambda_L$ satisfies the Bernoulli distribution  for $0< p <1$
 \begin{equation}
 p(r_x) = p \delta(r_x-1) + (1-p) \delta (r_x).
 \end{equation}
 
 \subsection{The Gibbs state}
Here, we define Gibbs state for the Hamiltonian.
The  partition function as a function
 of  $(\beta, h) \in [0,\infty) \times {\mathbb R}$ and a sequence $r = (r_i)_{i\in C_L}$ is defined by
\begin{equation}
Z_L(\beta, h,  r) :=  \sum_{\sigma\in \Sigma_L} e^{ - \beta H(r,\sigma)}.
\end{equation}
The  expectation of a function of spin configuration $f(\sigma)$ in the Gibbs state is given by
\begin{equation}
\langle f(\sigma) \rangle =\frac{1}{Z_L(\beta,h,r)} \sum_{\sigma\in \Sigma_L} f(\sigma)  e^{ - \beta H(r,\sigma)}.
\end{equation}
Define a function of  $\beta \in [0,\infty)$ and a sequence $J = (r_i)_{i \in C_L}$ by
\begin{equation}
\psi_L(\beta,r) := \frac{1}{|\Lambda_L|} \log Z_L(\beta,h,r), \\
\end{equation}
$-\frac{|\Lambda_L|}{\beta}\psi_L(\beta,r)$ is called free energy in statistical physics.
The following  function $p_L:[0,\infty) \rightarrow {\mathbb R}$ is defined by
the expectation of $\psi_L(\beta,h,r)$  over $r$
\begin{eqnarray}
p_L(\beta,h,v):={\mathbb E} \psi_L(\beta,h,r) ,
\end{eqnarray}
where ${\mathbb E}$ denotes the expectation over the i.i.d. random variables $r$ with a variance $v$.

\subsection{Replica symmetry}
Next, we explain
replica symmetry breaking phenomena
which apparently violate self-averaging
of the overlap between two replicated quantities in a replica symmetric expectation.
Let  $\sigma^a (a=1, \cdots, n)$ be $n$ replicated  spin configurations, and consider the following Hamiltonian
$$
H(r,\sigma^1, \cdots, \sigma^n):= \sum_{a=1}^n H(r,\sigma^a),
$$
where replicated spin configurations share the same quenched randomness $r$.
This Hamiltonian is invariant under an arbitrary permutation $s (\in S_n)$,  which is a bijection acting on $\{1,2, \cdots, n\}$.
$$ H(r, \sigma^1, \cdots, \sigma^n)=H(r,\sigma^{s 1}, \cdots, \sigma^{s n} )$$

This permutation symmetry is the replica symmetry.
The spin overlap $R_{a,b}$ between two replicated spin configurations is defined by
\begin{equation}
R_{a,b}:=\frac{1}{|\Lambda_L|}\sum_{x \in \Lambda_L} \sigma_x^a \sigma_x^b.
\label{Rdef}
\end{equation}
for all models except for  the site-diluted Ising model. In this exceptional model,  the spin overlap is defined by  
\begin{equation}
R_{a,b} := \frac{1}{|\Lambda_L|}\sum_{x \in \Lambda_L}r_x^2 \sigma_x^a \sigma_x^b.
 \label{sdiRdef}
\end{equation}
If the distribution of  the overlap has a finite variance, it implies the replica symmetry breaking as observed  in the Sherrington-Kirkpatrick model \cite{Pr,T2,T}.
In Chatterjee's definition \cite{C2}, we say that the replica symmetry breaking occurs  if the finite  variance  calculated in the replica symmetric expectation
in the infinite-volume limit
$$
\lim_{L \rightarrow \infty} {\mathbb E}  \langle {\Delta R_{1,2} }^2 \rangle
 \neq 0,
$$
where
$\Delta R_{1,2} := R_{1,2} -{\mathbb E }  \langle R_{1,2} \rangle  $.
Chatterjee has given this definition of the replica symmetry breaking and
proven
\begin{equation}
\lim_{L \rightarrow \infty} {\mathbb E}  \langle{ \Delta R_{1,2} }^2 \rangle
 =0,
 \label{0Var}
 \end{equation}
in the Gaussian random field Ising model \cite{C2}. In the present paper, we extend his proof to non-Gaussian
disordered  Ising systems with the FKG property under a uniform field.

\subsection{Main theorem}

Consider a model satisfying the following  assumptions. \\

\noindent
{\it {\bf Assumption 1}   The infinite-volume limit \begin{equation}
\lim_{L\to\infty}p_L(\beta,h) =p(\beta,h),
\end{equation}
exists for each $\beta \in [0,\infty) \times {\mathbb R}$.
}\\

\noindent
{\it {\bf Assumption 2}   Let $f,g$ be monotonically increasing function of spin configuration $\sigma \in \Sigma_L$.
The FKG inequality  \cite{FKG}
\begin{equation}
\langle f(\sigma) ;g(\sigma) \rangle:=\langle f(\sigma) g(\sigma) \rangle -\langle f(\sigma) \rangle \langle g(\sigma) \rangle \geq 0,
\end{equation}
is valid for any $(\beta, h) \in [0,\infty) \times {\mathbb  R}$
and for an arbitrarily fixed random sequence  $r$. 
}\\

\noindent{\it {\bf Assumption 3}   
The expectation and the variance  of random variables $J_X(r)$  for  $X \in B_L$  and $h_x(r)$  for $x \in \Lambda_L$
exist as a  finite value.}

Note that Assumption 1 is proven for a class of models with random short-range  interactions \cite{C2,CL}.
Under these Assumptions, we prove the following theorem.

{\theorem In a model defined by the Hamiltonian (\ref{hamil}) satisfying Assumptions 1, 2 and 3
the following  variance vanishes \label{MT}
\begin{equation}
\lim_{L \rightarrow \infty} [
{\mathbb E}  \langle {R_{1,2} }^2 \rangle - ({\mathbb E }  \langle R_{1,2} \rangle  )^2 ]=0,
\end{equation}
for almost all 
$(\beta,h) \in [0,\infty) \times {\mathbb R}$
 in the infinite-volume limit.
}
\section{Proof}
\subsection{Variance inequalities}

To prove Theorem \ref{MT}, we introduce an artificial Gaussian random field perturbing the original Hamiltonian.
{\definition Let $g=(g_x)_{x \in \Lambda_L}$ be a sequence of   i.i.d. standard Gaussian random variables and 
define a function of  $g$ and $\sigma \in \Sigma_L$ by
\begin{equation}
\xi_L(g, \sigma) :=\frac{1}{|\Lambda_L|} \sum_{x\in \Lambda_L} g_x \sigma_x.
\end{equation}
For the exceptional case in the site-diluted Ising model,
\begin{equation}
\xi_L(g,\sigma) :=\frac{1}{|\Lambda_L|} \sum_{x\in \Lambda_L} g_x r_x \sigma_x.
\label{xisdi}
\end{equation}
Define the following  perturbed  Hamiltonian 
\begin{equation}
H_\mu (q,\sigma) := H(r,\sigma) -\mu |\Lambda_L|\xi_L(g,\sigma) ,
\label{perthamil}
\end{equation}
with  a coupling constant $\mu \in {\mathbb R}$. 
Let  $q=(q_i)_{i=1,2, \cdots, N}:=(g_x, r_i)_{x\in \Lambda_L, i \in C_L}$  be  a combined sequence of random variables 
labeled by numbers $i=1,\cdots, N$,  and  ${\mathbb E}$ denotes the expectation over all random variables $(q_i)_{i=1,\cdots, N}$.
Define a partition function of the perturbed Hamiltonian 
\begin{equation}
Z_L(\beta, h, \mu, q):=\sum_{\sigma \in \Sigma_L} e^{-\beta H_\mu}.
\end{equation}
Define the corresponding functions 
\begin{eqnarray}
&& \psi_L(\beta, h,\mu,  q) := \frac{1}{|\Lambda_L|} \log Z_L(\beta, h, \mu, q), \\ 
&&p_L(\beta, h, \mu) := {\mathbb E} \psi_L(\beta, h,\mu,  q), \\ 
&&p(\beta, h, \mu) := \lim_{L\to \infty} p_L(\beta,h,\mu).
\end{eqnarray}
And denote the expectation of an arbitrary function $f(\sigma)$ of spin configuration $\sigma \in  \Sigma_L$
in the perturbed Gibbs state with this Hamiltonian by
\begin{equation}
\langle f(\sigma) \rangle_\mu:= \frac{1}{Z_L(\beta, h, \mu, q)}\sum_{\sigma \in \Sigma_L} f(\sigma) e^{-\beta H_\mu(q,\sigma)}
 \end{equation}
}
Note that this model defined by the above Hamiltonian satisfies the FKG property still.

 {\lemma \label{varpsi} 
There exists a positive number $C$ independent of the system size $L$, such that  the variance of $\psi_L$ is bounded by 
\begin{equation}
{\mathbb E} \psi_L(\beta, h,\mu,  q) ^2- p_L(\beta, h, \mu)^2 \leq \frac{C}{|\Lambda_L|},
 \end{equation}
 for any $(\beta, h, \mu)\in (0,\infty) \times {\mathbb R}^2.$\\
 Proof.}  
 For an integer  $m = 1, 2, \cdots, N$  define a symbol 
${\mathbb E}_m$ which denotes the expectation over 
random variables  $( q_j)_{j > m}$. Note that ${\mathbb E}_0 ={\mathbb E}$ is the expectation over the all random variables $q=(q_j)_{j=1,2, \cdots, N} $,
and ${\mathbb E}_N$ is the identity mapping.

Here, we represent $\psi_L(h)$ as a function of a sequence of random variables $q=(q_j)_{j =1, \cdots, N}$ for lighter notation. 
 \begin{eqnarray}
&& {\mathbb E} \psi_L(q)^2- ({\mathbb E} \psi_L(q))^2\\&&= {\mathbb E} ({\mathbb E}_N \psi_L(q) )^2
- {\mathbb E} ({\mathbb E}_0 \psi_L(q))^2\\
&&=\sum_{m=1} ^N{\mathbb E}[ ({\mathbb E}_m \psi_L( q) )^2
- ({\mathbb E}_{m-1} \psi_L( q))^2].
 \end{eqnarray}
 In the $m$-th term, regard $\psi_L (q_m)$ as a function of $q_m$. Let $q'_m$ be an independent random variable satisfying the same distribution as that of $q_m$, and
${\mathbb E}'$ denotes  an expectation over only $q'_m$.  Note that
$
{\mathbb E}_{m-1} \psi_L(q_1, \cdots, q_m, \cdots,  q_N) = {\mathbb E}_{m} {\mathbb E}' \psi_L(q_1, \cdots, q_m', \cdots,  q_N).
 $
Let $C_m \subset C_L$ be a collection of all $X \in C_L$ such that  $J_{X} (q)$ depends on $q_m$. 
 \begin{eqnarray}
&&{\mathbb E}[ ({\mathbb E}_m \psi_L(q_m ))^2
- ({\mathbb E}_{m-1} \psi_L(q_m))^2] \\
&&
={\mathbb E} [({\mathbb E}_m \psi_L( q_m ))^2
- ({\mathbb E}_{m}{\mathbb E}' \psi_L( q'_m))^2] \\&&
=  {\mathbb E}
[ {\mathbb E}_m( \psi_L(q_m) -  {{\mathbb E}}' \psi_L( q_m') ) ]^2\\
&&={\mathbb E} [ {\mathbb E}_m {{\mathbb E}}'( \psi_L( q_m ) -   \psi_L( q_m')) ]^2\\&&
={\mathbb E} [{\mathbb E}_m  {{\mathbb E}}'  \psi_L( q_m) -\psi_L( q_m') ]^2,\\&&
={\mathbb E} \Big[{\mathbb E}_m  {{\mathbb E}}' \int_{q'_m} ^{q_m }ds \frac{\partial}{\partial s} \psi_L( s)  \Big]^2,\\&&
=\frac{1}{|\Lambda_L|^2}{\mathbb E} \Big[{\mathbb E}_m  {{\mathbb E}}' \int_{q'_m} ^{q_m }ds \beta  \sum_{X \in C_m }  \frac{\partial J_{X} (s) }{\partial s}\langle \sigma_{X }\rangle_{ s} \Big]^2\\&&
=\frac{1}{|\Lambda_L|^2}{\mathbb E} \Big[{\mathbb E}_m  {{\mathbb E}}' \beta  \sum_{X \in C_m }  \int_{J_{X}(q'_m)} ^{J_{X}(q_m) }d J_{X} 
 \langle \sigma_{X }\rangle_{s} \Big]^2\\&&
\leq \frac{1}{|\Lambda_L|^2}{\mathbb E}  [{{\mathbb E}}'  \beta \sum_{X\in C_m} |J_{X}(q'_m)-J_{X}(q_m)| ]^2  
\leq \frac{C'}{N^2}
 \end{eqnarray}
 where $C'$ is a positive number independent of $N$, and
 we denote the Gibbs expectation  in the conditional probability under $ q_m$ by
 $$ \langle f(\sigma) \rangle_{q_m }= \frac{1}{Z_L( q_m)} \sum_{\sigma\in\Sigma_L}  f(\sigma) e^{-\beta H (q_m,\sigma)}.
 $$
 Since $|\Lambda_L| \leq |C_L| \leq C_1|\Lambda_L|$,   we have $2 |\Lambda_L|  \leq N \leq (C_1+1)|\Lambda_L|$ and  therefore
\begin{equation}
{\mathbb E} \psi_L(q)^2- ({\mathbb E} \psi_L( q))^2 \leq \frac{C'}{2|\Lambda_L|}.
\end{equation}
This completes the proof. $\Box$

To show the existence of the infinite-volume limit of the expectation of the overlap,
we use the standard convexity argument \cite{T}.
{\lemma \label{exR}
 For almost all $\mu \in {\mathbb R}\setminus\{0\}$, the following  expectation of the 
overlap exists  in the infinite-volume limit, and it is represented in terms of the derivative of $p(\beta, h, \mu) $
\begin{equation}
\lim_{L\to \infty }{\mathbb E}\langle R_{1,2}\rangle_\mu =1- \frac{1}{\beta^2\mu} \frac{\partial p}{\partial \mu}.
\end{equation}
Proof. } 
Here we regard  $p_L(\mu)$ and $p(\mu)$ as functions of $\mu$. 
Define a function $e_L(\epsilon, \mu)$ of $\epsilon > 0$ and $\mu \in {\mathbb R}$
\begin{eqnarray} e_L(\epsilon, \mu):=\frac{1}{\epsilon}[|p_L(\mu+\epsilon )-p({\mu}+\epsilon)|+|p_L({\mu}- \epsilon)-p({\mu}-\epsilon)|
+|p_L({\mu} )-p({\mu})|].
\end{eqnarray}
Since $p_L$ and $p$ are convex functions of $\mu$, we have
\begin{eqnarray}
&&\frac{\partial p_L}{\partial \mu}(\mu) - \frac{\partial  p}{\partial \mu}(\mu)
\leq \frac{1}{\epsilon} [p_L(\mu+\epsilon)-p_L(\mu)]- \frac{\partial  p}{\partial \mu}\nonumber \\
&&\leq \frac{1}{\epsilon} [p_L(\mu+\epsilon)-p_L(\mu)-p(\mu+\epsilon)+p(\mu+\epsilon)
+p(\mu)-p(\mu) ]- \frac{\partial  p}{\partial \mu}(\mu) \nonumber \\
&&\leq \frac{1}{\epsilon} [ |p_L(\mu+\epsilon)-p(\mu+\epsilon)|
+|p(\mu)-p_L(\mu)| +\frac{1}{\epsilon}[ p(\mu+\epsilon)-p(\mu)] - \frac{\partial  p}{\partial \mu}(\mu) \nonumber \\
&&\leq e_L(\epsilon,\mu)
+  \frac{\partial p}{\partial \mu}(\mu+\epsilon) - \frac{\partial  p}{\partial \mu}(\mu). \nonumber
\end{eqnarray}
As in the same calculation, we have
\begin{eqnarray}
&&\frac{\partial p_L}{\partial \mu}(\mu) - \frac{\partial  p}{\partial \mu}(\mu)
\geq \frac{1}{\epsilon}[p_L(\mu)-p_L(\mu-\epsilon)] - \frac{\partial  p}{\partial \mu}(\mu) \nonumber \\&&
\geq -e_L(\epsilon,\mu) + \frac{\partial p}{\partial \mu}(\mu-\epsilon)- \frac{\partial  p}{\partial \mu}(\mu) . \nonumber
\end{eqnarray}
Both inequalities imply
\begin{eqnarray}
\Big|\frac{\partial p_L}{\partial \mu}(\mu) - \frac{\partial p}{\partial \mu}(\mu)\Big| \leq
 e_L(\epsilon,\mu)
+  \frac{\partial p}{\partial \mu}(\mu+\epsilon) -  \frac{\partial p}{\partial \mu}(\mu-\epsilon).
\end{eqnarray}
In the infinite-volume limit, $e_L(\epsilon, \mu)$ vanishes for an arbitrary $\epsilon >0$ by Assumption 1,
\begin{eqnarray}
&&\lim_{L\to\infty} \Big|\beta^2\mu\Big(1-\frac{1}{|\Lambda_L|}\sum_{x\in \Lambda_L}   {\mathbb E} \langle \sigma_x \rangle_\mu^2\Big)- \frac{\partial p}{\partial \mu}(\mu)\Big| \\
&&=\lim_{L\to\infty} \Big|\beta {\mathbb E} \langle \xi_L\rangle_\mu- \frac{\partial p}{\partial \mu}(\mu)\Big|=
\lim_{L\to\infty} \Big|\frac{\partial p_L}{\partial \mu}(\mu) - \frac{\partial p}{\partial \mu}(\mu)\Big| \leq
 \frac{\partial p}{\partial \mu}(\mu+\epsilon) -  \frac{\partial p}{\partial \mu}(\mu-\epsilon).
 \end{eqnarray}
 Since the convex function $p(\mu)$ is continuously differentiable almost everywhere and $\epsilon >0$ is arbitrary, 
 this limit vanishes for almost all $\mu \in {\mathbb R}\setminus\{0\}.$
 This completes the proof. 
$\Box$

To show the following bound on the Hamiltonian density, 
we use the standard convexity argument also \cite{T}. 
{\lemma  \label{Delta} 
 For  almost all $\mu \in {\mathbb R}$,  the following infinite-volume limit vanishes 
 \begin{eqnarray}
\lim_{L\to\infty} {\mathbb E} |\langle \xi_L \rangle_\mu -{\mathbb E} \langle \xi_L \rangle_\mu| =0.
\end{eqnarray}
Proof.
}Here we regard  $\psi_L(\mu)$, $p_L(\mu)$ and $p(\mu)$ as functions of $\mu$. 
Define a function $w_L(\epsilon, \mu)$ of $\epsilon > 0$ and $\mu$.
\begin{eqnarray}
w_L(\epsilon,\mu)& :=& \frac{|\psi_L(\mu+\epsilon )-p_L(\mu+\epsilon)|}{\epsilon}
+\frac{|\psi_L(\mu)-p_L(\mu)|}{\epsilon} \nonumber \\
&+&\frac{|\psi_L(\mu- \epsilon)-p_L(\mu-\epsilon)|}{\epsilon}.
\end{eqnarray}
 Lemma  \ref{varpsi} implies
\begin{eqnarray}
&&{\mathbb E}w_L(\epsilon,\mu) \\
&&\leq \frac{1}{\epsilon}\sqrt{ {\mathbb E}[\psi_L(\mu+\epsilon)-p_L(\mu+\epsilon)]^2}  + \frac{1}{\epsilon}\sqrt{ {\mathbb E}[\psi_L(\mu)-p_L(\mu)]^2} 
\\&&+
\frac{1}{\epsilon}\sqrt{ {\mathbb E}[\psi_L(\mu-\epsilon)-p_L(\mu-\epsilon)]^2} \\ 
&&\leq \frac{3}{\epsilon}\sqrt{\frac{C}{ |\Lambda_L|}} \label{boundw}
\end{eqnarray}
Since $\psi_L$ and $p_L$ are convex functions of $\mu$, we have
\begin{eqnarray}
&&\frac{\partial \psi_L}{\partial \mu}(\mu) - \frac{\partial  p_L}{\partial \mu}(\mu)
\leq \frac{1}{\epsilon} [\psi_L(\mu+\epsilon)-\psi_L(\mu)]- \frac{\partial  p_L}{\partial \mu}\nonumber \\
&&\leq \frac{1}{\epsilon} [\psi_L(\mu+\epsilon)-p_L(\mu+\epsilon)+p_L(\mu+\epsilon)-p_L(\mu)
+p_L(\mu)-\psi_L(\mu) ]- \frac{\partial  p}{\partial \mu}(\mu) \nonumber \\
&&\leq \frac{1}{\epsilon} [ |\psi_L(\mu+\epsilon)-p_L(\mu+\epsilon)|
+|p_L(\mu)-\psi_L(\mu)| +\frac{1}{\epsilon}[ p_L(\mu+\epsilon)-p_L(\mu)] - \frac{\partial  p_L}{\partial \mu}(\mu) \nonumber \\
&&\leq w_L(\epsilon,\mu)
+  \frac{\partial p_L}{\partial \mu}(\mu+\epsilon) - \frac{\partial  p_L}{\partial \mu}(\mu). \nonumber
\end{eqnarray}
As in the same calculation, we have
\begin{eqnarray}
&&\frac{\partial \psi_L}{\partial \mu}(\mu) - \frac{\partial  p_L}{\partial \mu}(\mu)
\geq \frac{1}{\epsilon}[\psi_L(\mu)-\psi_L(\mu-\epsilon)] - \frac{\partial  p_L}{\partial \mu}(\mu) \nonumber \\&&
\geq -w_L(\epsilon,\mu) + \frac{\partial p_L}{\partial \mu}(\mu-\epsilon)- \frac{\partial  p_L}{\partial \mu}(\mu) . \nonumber
\end{eqnarray}
Both inequalities imply
\begin{eqnarray}
\Big|\frac{\partial \psi_L}{\partial \mu}(\mu) - \frac{\partial p_L}{\partial \mu}(\mu)\Big| \leq
 w_L(\epsilon,\mu)
+  \frac{\partial p_L}{\partial \mu}(\mu+\epsilon) -  \frac{\partial p_L}{\partial \mu}(\mu-\epsilon).
\end{eqnarray}
Then, we have
\begin{eqnarray}
&&{\mathbb E} \Big|\frac{\partial \psi_L}{\partial \mu}(\mu) - \frac{\partial p_L}{\partial \mu}(\mu)\Big| \leq
{\mathbb E} w_L(\epsilon,\mu)
+  \frac{\partial p_L}{\partial \mu}(\mu+\epsilon) -  \frac{\partial p_L}{\partial \mu}(\mu-\epsilon)
\\
&&\leq \frac{3}{\epsilon}\sqrt{\frac{C}{ |\Lambda_L|}}+  \frac{\partial p_L}{\partial \mu}(\mu+\epsilon) -  \frac{\partial p_L}{\partial \mu}(\mu-\epsilon).
\end{eqnarray}
This and the bound 
 (\ref{boundw}) imply
\begin{eqnarray}
&&\beta\int_{\mu_1}^{\mu_2} d\mu {\mathbb E} |\langle \xi_L \rangle_\mu -{\mathbb E} \langle \xi_L \rangle_\mu | \\
&&\leq \frac{3(\mu_2-\mu_1)}{\epsilon}\sqrt{\frac{C}{ |\Lambda_L|}}+ \int_{\mu_2-\epsilon} ^{\mu_2+\epsilon} d\mu p_L'(\mu) 
- \int_{\mu_1-\epsilon} ^{\mu_1+\epsilon} d\mu p_L'(\mu)\\
&&=
 \frac{3(\mu_2-\mu_1)}{\epsilon}\sqrt{\frac{C}{ |\Lambda_L|}}+ \int_{\mu_2-\epsilon} ^{\mu_2+\epsilon} d\mu 
\frac{1}{ |\Lambda_L|}\sum_{x\in \Lambda_L} {\mathbb E}g_x \langle \sigma_x\rangle_\mu 
\\&&- \int_{\mu_1-\epsilon} ^{\mu_1+\epsilon} d\mu \frac{1}{|\Lambda_L|}\sum_{x\in \Lambda_L} {\mathbb E}g_x \langle \sigma_x \rangle_\mu\\
&&=
 \frac{3(\mu_2-\mu_1)}{\epsilon}\sqrt{\frac{C}{ |\Lambda_L|}}+ \int_{\mu_2-\epsilon} ^{\mu_2+\epsilon} d\mu 
\frac{\beta\mu}{ |\Lambda_L|}\sum_{x\in \Lambda_L} {\mathbb E}(1- \langle \sigma_x\rangle_\mu ^2)
\\&&- \int_{\mu_1-\epsilon} ^{\mu_1+\epsilon} d\mu \frac{\beta\mu}{|\Lambda_L|}\sum_{x\in \Lambda_L} {\mathbb E}(1- \langle \sigma_x \rangle_\mu^2)\\
&&\leq  \frac{3(\mu_2-\mu_1)}{\epsilon}\sqrt{\frac{C}{ |\Lambda_L|}}+2\beta(|\mu_2|+|\mu_1|) \epsilon.
\end{eqnarray}
Since the above bound is valid for an arbitrary $\epsilon>0$, the right-hand side is bounded by $|\Lambda_L|^{-\frac{1}{4}}$
 times a positive number for $\epsilon= |\Lambda_L|^{-\frac{1}{4}}$. Thus,  the following  infinite-volume limit vanishes
\begin{equation}
\lim_{L\to\infty}\int_{\mu_1}^{\mu_2} d\mu {\mathbb E} |\langle \xi_L \rangle_\mu -{\mathbb E} \langle \xi_L \rangle_\mu | =0.
\end{equation}
The integrand vanishes for 
almost all $\mu\in{\mathbb R}$, since  the integration  interval $(\mu_1,\mu_2)$ is arbitrary and the integrand 
 has a uniform bound 
\begin{eqnarray}
&&{\mathbb E} |\langle \xi_L \rangle_\mu -{\mathbb E} \langle \xi_L \rangle_\mu | 
\leq \sqrt{ {\mathbb E} (\langle \xi_L \rangle_\mu - {\mathbb E} \langle \xi_L \rangle_\mu )^2 } = \sqrt{ {\mathbb E} \langle \xi_L \rangle_\mu^2 -( {\mathbb E} \langle \xi_L \rangle_\mu )^2 } \\
&&=\sqrt{ \frac{1}{|\Lambda_L|^2} \sum_{x,y\in \Lambda_L}[ {\mathbb E}g_xg_y \langle \sigma_x \rangle_\mu\langle \sigma_y \rangle_\mu  -  
{\mathbb E} g_x \langle \sigma_x \rangle_\mu {\mathbb E} g_y \langle \sigma_y \rangle_\mu ]} 
\\
&&=\sqrt{ \frac{1}{|\Lambda_L|^2} \sum_{x,y\in \Lambda_L}\Big[ {\mathbb E} \Big( \frac{\partial^2 }{\partial g_x \partial g_y}+ \delta_{x,y} \Big)
 \langle \sigma_x \rangle_\mu\langle \sigma_y \rangle_\mu  -  
{\mathbb E} \frac{\partial \langle \sigma_x \rangle_\mu}{\partial g_x} {\mathbb E} \frac{\partial\langle \sigma_y \rangle_\mu}{\partial g_y} \Big]} 
\\
&&
=\sqrt{ \frac{1}{|\Lambda_L|^2} \sum_{x,y\in \Lambda_L} {\mathbb E} [\beta^2 \mu^2\langle \sigma_x ;\sigma_y\rangle_\mu(
\langle \sigma_x \sigma_y\rangle_\mu-5 \langle \sigma_x \rangle_\mu\langle \sigma_y \rangle_\mu
) +\delta_{x,y} \langle \sigma_x\rangle_\mu^2]}\\
&&\leq \sqrt{12\beta^2\mu^2 +|\Lambda_L|^{-1}}\leq \sqrt{12\beta^2\mu^2 +1}.
 \end{eqnarray}
Also in the exceptional case  defined by (\ref{xisdi}), this is proven in the same way.  This completes the proof.
$\Box$\\

{\lemma \label{deltaR}
The variance of the overlap in the Gibbs state vanishes for any $\mu\in {\mathbb R}$ for almost all $h \in {\mathbb R}$ 
for an arbitrary sequence of  random variables $q=(q_i)_{i=1,\cdots, N}$ for any $\beta \in [0,\infty) $
\begin{equation}
\lim_{L \rightarrow \infty} [ \langle {R_{1,2} }^2 \rangle_\mu -  \langle R_{1,2} \rangle_\mu  ^2 ]=0.
\end{equation}
Proof. }
The following variance of the overlap defined by (\ref{Rdef}) is evaluated in terms of correlation function
\begin{eqnarray}
&&\langle R_{1,2} ^2\rangle_\mu-\langle R_{1,2} \rangle_\mu^2 =
\frac{1}{|\Lambda_L|^2}\sum_{x,y\in \Lambda} (\langle \sigma_x \sigma_y\rangle_\mu^2-\langle \sigma_x \rangle_\mu^2\langle  \sigma_y\rangle_\mu^2) \\
&& = \frac{1}{|\Lambda_L|^2}\sum_{x,y\in \Lambda} (\langle \sigma_x \sigma_y\rangle_\mu-\langle \sigma_x \rangle_\mu\langle  \sigma_y\rangle_\mu)(\langle \sigma_x \sigma_y\rangle_\mu +\langle \sigma_x \rangle_\mu\langle  \sigma_y\rangle_\mu) \\
&&\leq  \frac{1}{|\Lambda_L|^2}\sum_{x,y\in \Lambda} |\langle \sigma_x \sigma_y\rangle_\mu-\langle \sigma_x \rangle_\mu\langle  \sigma_y\rangle_\mu||\langle \sigma_x \sigma_y\rangle_\mu +\langle \sigma_x \rangle_\mu\langle  \sigma_y\rangle_\mu|
\\
&&\leq  \frac{2}{|\Lambda_L|^2}\sum_{x,y\in \Lambda} |\langle \sigma_x \sigma_y\rangle_\mu-\langle \sigma_x \rangle_\mu\langle  \sigma_y\rangle_\mu|\\
&&= \frac{2}{|\Lambda_L|^2}\sum_{x,y\in \Lambda} (\langle \sigma_x \sigma_y\rangle_\mu-\langle \sigma_x \rangle_\mu\langle  \sigma_y\rangle_\mu)\\
&&=\frac{2}{|\Lambda_L|^2 \beta}\sum_{x\in \Lambda} \frac{\partial}{\partial h}\langle \sigma_x \rangle_\mu.
\end{eqnarray}
 The FKG inequality has been used.
Therefore, the integralation over $h \in (h_1, h_2) $ in an arbitrary interval becomes
\begin{eqnarray}
\int_{h_1}^{h_2} dh (\langle R_{1,2}^2\rangle_{\mu,h} -\langle R_{1,2} \rangle_{\mu,h}^2) \leq \frac{2}{\beta|\Lambda_L|^2 }\sum_{x\in \Lambda} [
\langle \sigma_x \rangle_{\mu,h=h_2} -\langle \sigma_x \rangle_{\mu, h=h_1}]  \leq \frac{4}{ \beta|\Lambda_L|},
\end{eqnarray}
where the dependence of the Gibbs expectation in the  uniform field $h$ has been denoted explicitly.
The infinite-volume limit of the variance of the overlap vanishes for almost all $h \in {\mathbb R}$ for any $(q_i)_{i=1, \cdots, N}$.
Also in the exceptional case, the same way proves that  the variance the overlap defined by (\ref{sdiRdef}) vanishes.
$\Box$

{\lemma  \label{deltah}The following variance vanishes in the infinite-volume limit %
\begin{equation}
\lim_{L\to\infty} 
 {\mathbb E}\langle \delta { \xi_L}^2\rangle_\mu =0 
 , 
\label{a1}
\end{equation}
for 
almost all 
$\mu \in {\mathbb R}$, where $\delta \xi_L:= \xi_L- \langle \xi_L \rangle_\mu$.\\
Proof.
}
The variance  ${\mathbb E} \langle \delta \xi_L^2 \rangle_\mu $  is represented in the derivative of the 
one point function
\begin{eqnarray}
&&
{\mathbb E} \langle \delta \xi_L^2\rangle_\mu =
\frac{1}{|\Lambda_L|^2} \sum_{x,y \in \Lambda}{ \mathbb E} g_x g_y (\langle \sigma_x \sigma_y \rangle_\mu-\langle \sigma_x \rangle_\mu \langle\sigma_y \rangle_\mu)\\&&
= 
\frac{1}{|\Lambda_L |^2\beta } \frac{\partial}{ \partial \mu}
\sum_{x \in \Lambda_L} {\mathbb E}  g_x \langle\sigma_x\rangle_\mu
\end{eqnarray}
Integration over $\mu \in (\mu_1,\mu_2)$  yields
\begin{eqnarray}
&&\int_{\mu_1}^{\mu_2} d\mu {\mathbb E} \langle \delta \xi_L^2\rangle_\mu  = \frac{1}{|\Lambda_L| ^2\beta } 
\sum_{x \in \Lambda_L} {\mathbb E}  g_x( \langle\sigma_x\rangle_{\mu_2}- \langle\sigma_x\rangle_{\mu_1})\\
&& = \frac{\beta }{|\Lambda_L| ^2\beta } 
\sum_{x \in \Lambda_L} {\mathbb E} (\mu_2-\mu_2 \langle\sigma_x\rangle_{\mu_2}^2-\mu_1 
 + \mu_1 \langle\sigma_x\rangle_{\mu_1}^2 ) \leq \frac{\beta(|\mu_2|+|\mu_1|)}{|\Lambda_L|\beta}.
\end{eqnarray}
The limit vanishes
\begin{equation}
\lim_{L\to\infty} \int_{\mu_1}^{\mu_2} d\mu {\mathbb E} \langle \delta \xi_L^2\rangle_\mu  =0.
\end{equation}
Since the integration interval is arbitrary and the integrand has a uniform bound, this completes the proof.
Also in the exceptional case defined by (\ref{xisdi}), the same way proves that the variance of $\xi_L$ in the infinite-volume limit.
$\Box$

To show a bound on the following deviation of $\xi_L$ ,
  Lemma \ref{Delta} and Lemma \ref{deltah}  and.

{\lemma 
The following  limit vanishes
\begin{equation}
\lim_{L \rightarrow \infty} {\mathbb E} \langle|  {\Delta \xi_L} |\rangle_\mu=0,
\end{equation}
for almost all $\mu \in {\mathbb R}$, 
where $\Delta \xi_L := \xi_L - {\mathbb E} \langle \xi_L  \rangle_\mu$ \\
Proof.}
\begin{eqnarray}
&&{\mathbb E} \langle|  {\Delta \xi_L} |\rangle_\mu = {\mathbb E} \langle | \xi_L -  \langle \xi_L \rangle_\mu +
 \langle \xi_L  \rangle_\mu- {\mathbb E} \langle \xi_L \rangle |  \rangle_\mu  \\
&&\leq {\mathbb E} \langle | \delta \xi_L | \rangle_\mu + {\mathbb E}| \langle  \Delta \xi_L  \rangle_\mu |
\leq  \sqrt{{\mathbb E} \langle \delta  \xi_L^2  \rangle_\mu} + {\mathbb E} |\langle \Delta \xi_L \rangle_\mu|
\end{eqnarray}
Lemma \ref{Delta} and Lemma \ref{deltah}  imply
 \begin{equation}
\lim_{L\to\infty}\int_{\mu_1}^{\mu_2} d\mu {\mathbb E} \langle|\Delta \xi_L | \rangle_\mu =0.
\end{equation}
The integrand in the left hand side  vanishes in the infinite-volume limit for almost all $\mu \in {\mathbb R}$. 
This completes the proof. $\Box$

The Ghirlanda-Guerra identities are well-known as useful identities for disordered spin systems \cite{C1,C2,CG,CG2,CG3,GG,I2,T}.
 The following lemma gives the  Ghirlanda-Guerra identities for an arbitrary bounded function of spin configurations
  for almost all $\mu \in {\mathbb R}$.
{\lemma \label{GG} 
Let 
$f : \Sigma_L^n \to {\mathbb R}$ be a bounded measurable function of $n$ replicated spin configurations.
For almost all $\mu \in {\mathbb R}\setminus\{0\}$,
   the following identity is valid
in the infinite-volume limit of the perturbed model \begin{equation}
\lim_{L \rightarrow \infty} \Big[\sum_{a=2} ^n {\mathbb E}\langle R_{1,a} f\rangle_{\mu} -n {\mathbb E} \langle R_{1,n+1
} f \rangle_{\mu}
  + \ {\mathbb E}\langle R_{1,2} \rangle_{\mu}{\mathbb E}\langle f \rangle_{\mu}
\Big]=0.
\label{GGid}
\end{equation}
Proof.}
The following expectation has a bound
$$
|{\mathbb E}\langle  \Delta \xi_L  f  \rangle_{\mu} | \leq 
{\mathbb E}\langle | \Delta \xi_L||  f | \rangle_{\mu}  \leq  \sup_{{\bf \sigma} \in \Sigma_L^n} |f({\bf \sigma})| {\mathbb E}\langle | \Delta \xi_L| \rangle_{\mu}.
$$
Since $f$ is bounded, the limit vanishes 
$$
\lim_{L\to\infty}  |{\mathbb E}\langle  \Delta \xi_L  f  \rangle_{\mu}|  =0.
 $$
 The calculation of the left hand side gives
 $$
 {\mathbb E}\langle  \Delta \xi_L  f  \rangle_{\mu}= \beta \mu \Big[\sum_{a=2} ^n {\mathbb E}\langle R_{1,a} f\rangle_{\mu} -n {\mathbb E} \langle R_{1,n+1
} f \rangle_{\mu}
  + \ {\mathbb E}\langle R_{1,2} \rangle_{\mu}{\mathbb E}\langle f \rangle_{\mu}
\Big].
 $$
Since the left-hand side vanishes in the infinite-volume limit, the right-hand side also vanishes for  almost all $\mu \in {\mathbb R}\setminus\{0\}$.
This completes   the proof. 
$\Box$

Next, we prove the continuity of  ${\mathbb E} \langle R_{1,2} \rangle_\mu$  and  ${\mathbb E} \langle R_{1,2} \rangle_\mu^2$ at
$\mu=0$.
{\lemma \label{R}  
The following limit is identical to that at $\mu=0$
\begin{equation}
\lim_{\mu \to 0} \lim_{L\to\infty} {\mathbb E} \langle R_{1,2} \rangle_{\mu}=  \lim_{L\to\infty} {\mathbb E} \langle R_{1,2} \rangle_{0},
\end{equation}
for  almost all $h \in {\mathbb R}$ .\\
Proof.}  Represent the difference
\begin{eqnarray}
&&|\mathbb{E} \langle R_{1,2} \rangle_\mu -\mathbb{E} \langle R_{1,2} \rangle_0|
=\Big| \int_0^\mu ds \frac{\partial} {\partial s} \mathbb{E} \langle
  R_{1,2} \rangle_s \Big| 
  \\&& 
  = \frac{2 }{ |\Lambda_L|} \Big|\int_0^\mu ds
  \mathbb{E} \sum_{y\in\Lambda_L} \langle \sigma_y \rangle_s \frac{\partial}{\partial s} \langle \sigma_y \rangle_s \Big|\\
  && =\frac{2 \beta}{ |\Lambda_L|}\Big|\int_0^\mu ds \sum_{x,y\in\Lambda} \mathbb{E} g_x
  \langle \sigma_y \rangle_s(\langle \sigma_x  \sigma_y \rangle_s -\langle \sigma_x \rangle_s \langle \sigma_y \rangle_s) \Big| \\
  && =\frac{2 \beta}{ |\Lambda_L|}\Big| \int_0^\mu ds \sum_{x,y\in\Lambda} \mathbb{E} \frac{\partial}{\partial g_x}\langle \sigma_y \rangle_s
   (\langle \sigma_x  \sigma_y \rangle_s -\langle \sigma_x \rangle_s \langle \sigma_y \rangle_s)\Big| \\
  && = \frac{2 \beta^2}{ |\Lambda_L|}\Big| \int_0^\mu ds s \sum_{x,y\in\Lambda_L} \mathbb{E} (\langle \sigma_x  \sigma_y \rangle_s -\langle \sigma_x \rangle_s \langle \sigma_y \rangle_s) (\langle \sigma_x  \sigma_y \rangle_s - 3 \langle \sigma_x \rangle_s \langle \sigma_y \rangle_s) \Big|\\
  && \leq \frac{2 \beta^2}{ |\Lambda_L|} \int_0^\mu ds s \sum_{x,y\in\Lambda_L} \mathbb{E} |\langle \sigma_x  \sigma_y \rangle_s -\langle \sigma_x \rangle_s \langle \sigma_y \rangle_s |
  |\langle \sigma_x  \sigma_y \rangle_s - 3 \langle \sigma_x \rangle_s \langle \sigma_y \rangle_s| \\
  && \leq  \frac{8 \beta^2}{ |\Lambda_L|} \int_0^\mu ds s \sum_{x,y\in\Lambda_L} \mathbb{E} \langle \sigma_x ; \sigma_y \rangle_s 
  = \frac{8 \beta}{ |\Lambda_L|}  \int_0^\mu ds s
  \mathbb{E} \sum_{x\in\Lambda_L}  \frac{\partial}{\partial h} \langle \sigma_x \rangle_s
\end{eqnarray}
 The FKG inequality has been used.
Therefore, the integral over $h$ in an arbitrary interval becomes
\begin{eqnarray}
&&\int_{h_1}^{h_2} dh |{\mathbb E}\langle R_{1,2}\rangle_{\mu,h} -{\mathbb E}\langle R_{1,2} \rangle_{0,h}|
\leq \frac{8 \beta}{|\Lambda_L|}\int_0^\mu ds s
  \mathbb{E}\sum_{x\in \Lambda}
[\langle \sigma_x \rangle_{s,h=h_2} -\langle \sigma_x \rangle_{s,h=h_1}]\\
&&\leq \frac{8 \beta}{|\Lambda_L| }\int_0^\mu ds s
  \mathbb{E}\sum_{x\in \Lambda}2
 \leq 8 \beta \mu^2,
\end{eqnarray}
where we have represented the dependence of the Gibbs expectation in the  uniform field $h$ explicitly. 
Therefore, the integrand vanishes in the infinite-volume limit.
Also in the exceptional case  defined by (\ref{sdiRdef}), this is proven in the same way.
$\Box$

{\lemma \label{R2} 
The following limit is identical to that at $\mu=0$
\begin{equation}
\lim_{\mu\to 0}\lim_{L\to\infty} {\mathbb E} \langle R_{1,2} \rangle_{\mu}^2 = \lim_{L\to\infty} {\mathbb E} \langle R_{1,2} \rangle_{0}^2,
\end{equation}
for almost all $h \in {\mathbb R}$.
\\
Proof.} As in the proof of Lemma \ref{R},
\begin{eqnarray}
&&|\mathbb{E} \langle R_{1,2} \rangle_\mu^2 -\mathbb{E} \langle R_{1,2} \rangle_0^2|
 = \Big|  \int_0^\mu ds \frac{\partial} {\partial s} \mathbb{E} \langle R_{1,2} \rangle_s^2  \Big|\\
  && = 2 \Big|\int_0^\mu ds \mathbb{E} \langle R_{1,2} \rangle_s
  \frac{\partial}{\partial s} \langle R_{1,2} \rangle_s\Big| = 2\Big| \int_0^\mu ds \mathbb{E} \langle R_{1,2} \rangle_s \frac{\partial} {\partial s}
 |\Lambda_L|^{-1}\sum_{y\in\Lambda_L}  \langle \sigma_y \rangle_s^2 \Big| \\
  && = 4 |\Lambda_L|^{-1}\Big| \int_0^\mu ds \mathbb{E} \langle R_{1,2} \rangle_s\sum_{y\in\Lambda_L}  \langle \sigma_y \rangle_s \frac{\partial} {\partial s} \langle \sigma_y
  \rangle_s \Big|\\
&& =  4 \beta |\Lambda_L|^{ -1}\Big| \int_0^\mu ds \mathbb{E}
\sum_{x,y\in\Lambda} g_x \langle R_{1,2} \rangle_s \langle \sigma_y \rangle_s \langle \sigma_x ; \sigma_y \rangle_s\Big| \\
&& = 4 \beta |\Lambda_L|^{ -1}\Big| \int_0^\mu ds \mathbb{E}
\sum_{x,y\in\Lambda} \frac{\partial}{\partial g_x} \langle R_{1,2} \rangle_s \langle \sigma_y \rangle_s \langle \sigma_x ; \sigma_y \rangle_s\Big| \\
&& = 4 \beta |\Lambda_L|^{ -2} \Big|\int_0^\mu ds \mathbb{E}
\sum_{x,y,z\in\Lambda} (\langle \sigma_z \rangle_s^2
 \frac{\partial}{\partial g_x} \langle \sigma_y \rangle_s \langle \sigma_x ; \sigma_y \rangle_s + \langle \sigma_y \rangle_s \langle \sigma_x ; \sigma_y \rangle_s \frac{\partial}{\partial g_x} \langle \sigma_z \rangle_s^2 )\Big|\\
 && = 4 \beta^2 |\Lambda_L|^{ -2}\Big| \int_0^\mu ds s \mathbb{E} \sum_{x,y,z\in\Lambda} (  \langle \sigma_z \rangle_s^2 \langle \sigma_x ;\sigma_y \rangle_s (\langle \sigma_x  \sigma_y \rangle_s - 3 \langle \sigma_x \rangle_s \langle \sigma_y \rangle_s ) \\&&
 +2  \langle \sigma_z \rangle_s
 \langle \sigma_y \rangle_s \langle \sigma_x ; \sigma_y \rangle_s
 \langle \sigma_x ; \sigma_z \rangle_s )\Big|\\
 && \leq 4 \beta^2 |\Lambda_L|^{ -2} \int_0^\mu ds s \mathbb{E} \sum_{x,y,z\in\Lambda} (  \langle \sigma_z \rangle_s^2 |\langle \sigma_x ;\sigma_y \rangle_s| |\langle \sigma_x  \sigma_y \rangle_s - 3 \langle \sigma_x \rangle_s \langle \sigma_y \rangle_s| ) \\
 && +2  |\langle \sigma_z \rangle_s|
 |\langle \sigma_y \rangle_s| |\langle \sigma_x ; \sigma_y \rangle_s|
|\langle \sigma_x ; \sigma_z \rangle_s| )\\
&& \leq 4 \beta^2 |\Lambda_L|^{ -2} \int_0^\mu ds s \mathbb{E} \sum_{x,y,z\in\Lambda} (  |\langle \sigma_x ;\sigma_y \rangle_s| |\langle \sigma_x  \sigma_y \rangle_s - 3 \langle \sigma_x \rangle_s \langle \sigma_y \rangle_s|  \\
&& +2  |\langle \sigma_x ; \sigma_y \rangle_s|
|\langle \sigma_x ; \sigma_z \rangle_s| )
\leq 4 \beta^2 |\Lambda_L|^{-2} \int_0^\mu ds s \mathbb{E} \sum_{x,y,z\in\Lambda} (  4 \langle \sigma_x ;\sigma_y \rangle_s  + 4 \langle \sigma_x ; \sigma_y \rangle_s ) \\
&& = 32 \beta^2 |\Lambda_L|^{-1} \int_0^\mu ds s \mathbb{E} \sum_{x,y\in\Lambda_L}  \langle \sigma_x ; \sigma_y \rangle_s
 = 
32 \beta |\Lambda_L|^{ -1} \int_0^\mu ds s \mathbb{E}  \sum_{x\in\Lambda_L} \frac{\partial}{\partial h} \langle \sigma_x \rangle_s
\end{eqnarray}
 The FKG inequality has been used. Therefore, the integral over $h$ in an arbitrary interval becomes
\begin{eqnarray}
  && \int_{h_1}^{h_2} dh |{\mathbb E}\langle R_{1,2}\rangle_{\mu,h}^2 -{\mathbb E}\langle R_{1,2} \rangle_{0,h}^2| \\&&
   \leq 32 \beta |\Lambda_L|^{-1} \int_0^\mu ds s \mathbb{E}  \sum_{x\in\Lambda_L}
  [\langle \sigma_x \rangle_{\mu,h=h_2}-\langle \sigma_x \rangle_{\mu,h=h_1}] \\&&
   \leq 32 \beta |\Lambda_L|^{ -1} \int_0^\mu ds s \mathbb{E}  \sum_{x\in\Lambda_L}
2 \leq 32 \beta \mu^2,
\end{eqnarray}
where we have represented the dependence of the Gibbs expectation in the  uniform field $h$ explicitly.
Also in the exceptional case  defined by (\ref{sdiRdef}), this is proven in the same way.$\Box$

\subsection{Proof of Theorem \ref{MT}}
 First, let us remark the relation between three variances of the overlap. 
As discussed by Chatterjee \cite{C2},  the Ghirlanda-Guerra identities (\ref{GGid})
for $n=2, f=R_{1,2}$ and $n=3, f=R_{2,3}$ give the following relations
for almost all $\mu \in {\mathbb R}$. 
\begin{eqnarray}
&&3\lim_{L\to\infty}[{\mathbb E}  \langle {R_{1,2} }^2 \rangle_{\mu} -{\mathbb E}  \langle { R_{1,2} } \rangle_{\mu}^2]  \label{deltar}\\
=&&2\lim_{L\to\infty} [{\mathbb E}  \langle {R_{1,2} }^2 \rangle_{\mu} - ({\mathbb E}  \langle {R_{1,2} } \rangle_{\mu} )^2 ]
\\
=&&6\lim_{L\to\infty}[{\mathbb E}  \langle R_{1,2}  \rangle_{\mu}^2-({\mathbb E}  \langle { R_{1,2} } \rangle_{\mu})^2].
\end{eqnarray}
Since the first line (\ref{deltar}) vanishes from Lemma \ref{deltaR}, all lines vanishes for almost all $\mu \in {\mathbb R}$.
Therefore  Lemma \ref{R} and Lemma  \ref{R2} imply
\begin{equation}
\lim_{L\to\infty}{\mathbb E}  \langle R_{1,2}  \rangle_{0}^2=
\lim_{\mu\to0}\lim_{L\to\infty}{\mathbb E}  \langle R_{1,2}  \rangle_{\mu}^2=\lim_{\mu\to0}\lim_{L\to\infty}({\mathbb E}  \langle { R_{1,2} } \rangle_{\mu})^2
=\lim_{L\to\infty}({\mathbb E}  \langle { R_{1,2} } \rangle_{0})^2
\end{equation}
for almost all $h\in{\mathbb R}$.
 Since Lemma \ref{deltaR} implies 
 \begin{equation}
\lim_{L\to \infty}  ({\mathbb E}\langle R_{1,2} ^2\rangle_0  -{\mathbb E} \langle R_{1,2} \rangle_0 ^2) =0.
\end{equation} 
for almost all $h\in{\mathbb R}$,  also
 the  following variance of the overlap vanishes 
 $$\lim_{L\to\infty}[{\mathbb E}  \langle {R_{1,2} }^2 \rangle_{0}- ({\mathbb E}  \langle {R_{1,2} } \rangle_{0})^2 ]=0.$$
for almost all $h\in{\mathbb R}$. This completes the proof of Theorem \ref{MT}. $\Box$

\end{document}